\documentstyle [12pt,twoside]{article} 
  
\begin{document}
\centerline{\large{\bf General relativistic cosmology 
with no beginning of time}}
\vspace*{1.cm}
\centerline{\bf Redouane Fakir$^{\$}$}
\vspace*{0.5cm}
\centerline{\em Cosmology Group, Department of Physics and Astronomy}
\centerline{\em University of British Columbia}
\centerline{\em 6224 Agriculture Road, Vancouver, B.C. V6T 1Z1, Canada}

\vspace*{1.5cm}
\centerline{\bf Abstract} 
\vspace*{1.cm}
We find that general relativity can be naturally free of cosmological 
singularities. Several nonsingular models are currently available
that either assume ad hoc matter contents,
or are nonsingular only over a sector of solution space of zero measure,
or depart drastically from general relativity at high energies. 
After much uncertainty over 
whether cosmological inflation could help solve the initial-singularity 
problem, the prevailing belief today is that general relativistic cosmology,
with inflation or without,
is endemically singular. This belief was reinforced by recent singularity
theorems that take account specifically of inflation. 
Here, a viable inflationary cosmology is worked out that is naturally
free of singularities despite the fact that 1) it uses only classical general
relativity, 2) it assumes only the most generic inflationary matter contents,
and 3) it is a theory of the chaotic-inflation type. That type of inflation
is the most widely accepted today, as it demands 
the least fine-tuning of initial conditions. It is also shown how, by
dropping the usual simplification of minimal coupling between matter
and geometry, the null energy condition can be violated and the
relevant singularity theorems circumvented.

 \clearpage
The powerful singularity theorems of the sixties [1,2]
seemed to leave little room for avoiding a catastrophic
singularity in the cosmological past, implying a necessary primeval 
breakdown of the laws of physics. Naturally, questions were then
raised about the 
self-consistency of general relativity as a complete theory of gravity, a philosophical question the answer to which is still 
far from  clear [3].

There have been several successful attempts at building 
alternative theories of gravity that are nonsingular in the 
limit of very high energies (see, e.g., [4-6]). It has also been shown that
the equations of motion of several scalar-tensor theories of
gravity may admit nonsingular {\em asymptotic} solutions if the
scalar functions in the Lagrangian are reverse-engeneered
for that specific purpose [7-9]. Not surprisingly, most of the 
corresponding {\em full}
solutions are either highly unstable (and hence could not
be simulated numerically in the literature) or occupy a sector of 
measure zero in solution space or in parameter space.

Perhaps the most serious prospect of a nonsingular cosmology based
on ordinary gravity arose from the realization
that cosmological inflation [10] is probably future-eternal and
hence perhaps symmetrically past-eternal as well [11-16]. However, this
prospect was dimmed considerably by new inflation-specific
singularity theorems [17-21] that condemned a large class of inflationary 
models to be necessarily singular.

In hindsight, the failure of inflation to {\em guarantee} the avoidance
of a cosmological singularity is not surprising. The inflationary 
expansion itself, no matter how large, should not affect the global
topology of spacetime and hence should have no bearing on the 
existence or non-existence of global singularities. 
However, inflation {\em could} potentially
play a role in the singularity game by the mere fact that
it assumes the Universe to be dominated by scalar matter at very
high energies. Such a universe could easily violate the ``strong
energy condition'' which is assumed by several singularity 
theorems [2,22-24].

As we shall see below, it is the
{\em pre}-inflationary dynamics and energetics of the scalar matter
that are likely to help solve the initial-singularity problem. 
Once it enters the 
inflationary phase proper (``slow-roll'' conditions) the scalar field 
acquires dynamics and energetics that do seem to make the model fall under
the power of most singularity theorems. More specifically, it 
starts then to obey the ``null energy condition.'' 

We now discuss this latter point more quantitatively.
Then, we show in a specific model how this allows an inflationary
cosmology to be nonsingular, even though the characteristics of 
its ``inflationary stage'' proper may suggest otherwise. Finally,
we follow the evolution of this ordinary-gravity, chaotic inflationary
scenario starting from the late inflationary stage, when pre-galactic
density fluctuations are generated, back through the very brief but crucial
pre-inflationary phase, and finally through the cosmological bounce itself.
We confirm numerically that the model is a nonsingular cosmology
that is stable in solution space. We conclude with some additional
comments on the relevant recent literature.

To see how the quasi-de Sitter phase of inflation does not necessarily
effect the initial-singularity problem, consider first the  Raychaudhuri
equation for null geodesics in the absence of torsion and shear [2]:
\newline\begin{equation}
{d\theta\over dv} + {1\over 2}\theta^{2} = - R_{\mu\nu}N^{\mu}N^{\nu} \ \ .
\newline\end{equation}
$\theta$ is the divergence of the considered congruence of null
geodesics; $v$ is an affine parameter; $R_{\mu\nu}$ is the Ricci
curvature tensor; and the $N^{\mu}$ are components of a null $4$-vector
field that is tangent to a Robertson-Walker spacetime with metric
\newline\begin{equation}
ds^{2}=-dt^{2}+a^{2}(t)dS^{2} \;\; .
\end{equation}\newline
$dS^{2}$ is the metric of a homogeneous and isotropic $3$-space.
If one chooses the null $4$-vector field to be locally 
$(1,{\overrightarrow n}/a(t))$,
where ${\overrightarrow n}$ is a unit $3$-vector, one obtains
\newline\begin{equation}
{d\theta\over dv} + {1\over 2}\theta^{2} = 2 \left({{\ddot a}\over a} - 
{{\dot a}^{2}\over a^{2}} - {k\over a^{2}} \right) = 
2 \left({\dot H} - {k\over a^{2}} \right) \  \ ,
\newline\end{equation}
 where $H \equiv {\dot a}/a$ is the expansion rate of the spacetime
and $k = +1 \ , 0$ or $-1$ according to whether the spatial
$3$-geometry in Eq.(2) is closed, flat or open, respectively. Dots 
denote derivation with respect to $t$.
 
If one assumes general relativity, considers
a model that is driven by a scalar field $\phi$ with potential $V(\phi)$,
and makes the usual simplification of minimal coupling between $\phi$
and the scalar curvature ($\xi =0$ in Eq.(10) below),
then $\left( {\dot H} - k/a^{2} \right)$ is automatically negative definite:
\begin{equation}
{\dot H} - {k\over a^{2}} = - 4\pi G \ {\dot\phi}^{2} \ \ ,
\end{equation}
where $G$ is the gravitational constant.
This seemingly unavoidable fact causes enough past convergence through Eq.(3) that past null geodesics are
necessarily incomplete, i.e., the cosmology has a global past singularity
[17-21]. Note that a de-Sitter spacetime produces zero on the right-hand-side
of Eq.(4) only if it is {\em exactly} de-Sitter, 
in which case it cannot be part of a consistent cosmology.

The positivity of $-\left({\dot H} - k/a^{2} \right)$, which is essentially the sum of the density and the pressure of the model, 
also implies that the null energy condition
\begin{equation}
R_{\mu\nu}N^{\mu}N^{\nu} \ge 0  
\end{equation}
is satisfied, bringing such a cosmology under the wrath of several singularity
theorems [17-21]. We comment further on this and other energetic aspects at the end of this letter and in [25].

The key Eq.(4) above derives from a Lagrangian of the form
\newline\begin{equation}
{\cal L} = {R\over 16\pi G} + {1\over 2}\phi_{;\mu}\phi^{;\mu} - V(\phi) \ \ ,
\end{equation}\newline
where $R$ is the scalar curvature.
However, it is necessary for the renormalizability of the theory
that the Lagrangian contain an explicit coupling term between the 
scalar field and the scalar curvature [26-28]. The correct form of the  
Lagrangian is thus
\newline\begin{equation}
{\cal L} = {R\over 16\pi G} + {1\over 2}\xi\phi^{2}R + {1\over 2}\phi_{;\mu}\phi^{;\mu} - V(\phi) \ \ .
\end{equation}\newline
Note that in our conventions the conformal value of the nonminimal-coupling constant
$\xi$ is $-1/6$. Field theoretic considerations do not a priori constrain the value of $\xi$ (For a more complete discussion, see [29].) 
However, models with $\xi < 0$ are generally too pathological
to yield viable inflationary cosmologies [29,30-34]. In contrast, it was shown
some time ago that models with $\xi >0$, and especially those with $\xi >>1$,
lead to well behaved inflationary cosmologies with several attractive
features[30-37], such as the easing of the observational constraints on the 
constants that characterize $V(\phi)$.  

It can be shown (Eqs.(26,28) of [35]) that during the 
slow-roll phase of chaotic inflation
driven by the theory (7), one has 
\newline\begin{equation}
{\dot H} \approx - {8 H^{2}\over (1+6\xi) \phi^{2}} \ \ .
\end{equation}\newline

Hence, for the cosmologically interesting range $\xi > 0$, one has ${\dot H} < 0$, just as in Eq.(4) where the effect of nonminimal coupling was neglected. Thus, from the above discussion of the Raychaudhuri equation, one could presume that
the nonminimal-coupling adjustment would have no obvious effect on the 
singularity aspect of the model. However,  as we shall find soon,
the {\em pre}-inflationary stage of the model {\em is} drastically effected
by the nonminimal coupling, allowing e.g. ${\dot H}$ to assume positive
values at the earliest stages, and eventually leading to a nonsingular cosmological past.
 
Let us then analyze the full equations of motion [32-35] without assuming 
slow-roll conditions. The ``energy'' and ``momentum'' equations are,
respectively 
\newline\begin{equation}
3\left( H^{2} + {k\over a^{2}} \right) \left[ {1\over 8\pi G} + 
\xi\phi^{2} \right] =  {1\over 2}\dot{\phi}^{2} + V(\phi) 
- 6\xi H\phi\dot{\phi} \;\; ,
\end{equation}\newline
\newline\begin{equation}
\left( \dot{H} - {k\over a^{2}} \right) \left[ {1\over 8\pi G} + 
\xi\phi^{2} \right] = -{1\over 2}
\dot{\phi}^{2} + \xi H\phi\dot{\phi} - \xi\dot{\phi}^{2} 
- \xi\phi\ddot{\phi} \;\; .
\end{equation} 
The matter field evolves according to
\newline\begin{equation}
\ddot{\phi} + 3H\dot{\phi} - 6\xi\left( \dot{H} + 2H^{2} + 
{k\over a^{2}} \right) \phi + 
{\partial V(\phi)\over \partial \phi} = 0 \;\; .
\end{equation}\newline

At this point, one cannot limit oneself to looking for de Sitter-like asymptotic
solutions and, if there are any, conclude that the model must be 
nonsingular. In fact, many of the candidate nonsingular models in
the literature do have two exponentially expanding asymptotic solutions,
but fail to cross from one asymptotic region to the other,
or cross over unstable or zero-measure trajectories in phase space 
[8,9,38,25].
In fact, the present model illustrates well the caution that
must be exercised in inferring the existence or non-existence of singularities
from either the properties of the inflationary stage alone or those of the
asymptotic solutions in isolation. In the present case, for example, 
${\dot H} < 0$ during asymptotic slow-roll phases and yet the model
 can manage
to avoid crushing into a singularity by very briefly shifting 
away from the slow-roll conditions, allowing
${\dot H}$ to turn positive. Again, there clearly is not a causal relationship between the existence of de-Sitter like
asymptotic solutions and the avoidance of cosmological singularities.
This aspect is not always recognized, although it had become
clear during, e.g., the investigation of ``pre-big-bang'' cosmology 
[38-40,5].

Let us choose the potential to be of the simple and generic form
$V(\phi) = \lambda\phi^{4}$. We start at the late stage of inflation
when the slow-roll conditions hold. Then, the model evolves
according to 
\newline\begin{equation}
H^{2} \sim - B_{1} \log a + C_{1} \ \ , \ \ 
\phi^{2} \sim -B_{2} \log a + C_{2} \ \ , \ \
a \sim \exp{\left(C_{3}t-{1\over 4}B_{1}t^{2}\right)} \ \ ,
\end{equation}\newline
where $C_{1}$, $C_{2}$ and $C_{3}$
are constants that are determined by the initial conditions, while
$B_{1}$ and $B_{2}$ are {\em positive} constants that depend on the 
coupling parameters $\lambda$ and $\xi$:
\newline\begin{equation}
B_{2} = {16\over 1+6\xi} \ \ , \ \ B_{1} = {\lambda\over 3 \xi} B_{2} \ \ .
\end{equation}\newline
In addition, ${\dot\phi} \sim -B_{2}H/2\phi $ and fluctuations are
produced that will eventually result (when the relevant scales cross
back into the Hubble radius after the end of inflation) in pre-galactic
density perturbations with magnitude [35,36]
\newline\begin{equation}
{\delta\rho\over\rho} \propto {H^{2}\over {\dot\phi}(1+6\xi)^{1/2}} \ \ .
\end{equation}\newline

As announced above (see Eq.(8)), ${\dot H}$ is negative-definite during this 
relatively late stage of inflation: ${\dot H} \approx -B_{1}/2 $. As
one evolves the model towards the past, the slow-roll solution (12)
breaks down and, a priori, ${\dot H}$ can cover a wide range of 
values of either sign (see Eqs.(9-10)), depending on
the choice of parameters and initial conditions.
In a separate paper, we explore the many possible detailed behaviors
of the model that can result [25]. Our focus here is the possibility
of obtaining a complete and naturally nonsingular inflationary model
from fairly generic initial conditions. (By this, we mean of course
generic {\em inflationary} initial conditions. The question of whether
the latter are themselves probable among all possible initial conditions
is still open [43]).

By inspecting Eqs.(9-10) analytically for the behavior of the matter functions
($\phi$, ${\dot\phi}$, ${\ddot\phi}$) and the metric functions 
($a$, $H$, ${\dot H}$), 
we find that three types of scenarii are  possible. 

1) ${\dot H}$ remains negative-definite throughout and $H$ diverges at a finite past value of $t$ 
where $a(t)$ plunges to zero. I.e., one hits a singularity. This behavior
is generic in the case of an open spatial geometry ($k = -1$).

2) ${\dot H}$ first becomes positive (again, as we run the model toward
the past) but then decreases toward zero asymptotically. 
In this case, which obtains typically for spatially
flat geometries ($k=0$), $a(t)$ decreases asymptotically toward a finite minimal size. Such solutions are technically nonsingular, 
but are usually unstable for most combinations of parameters. 
Moreover, in the real Universe, the condition
of {\em rigorous} spatial flatness has arguably zero measure, so that unlike in
lower-energy cosmological problems, the cases of practical interest here
are probably only $k=-1$ and $k=+1$. 

3) ${\dot H}$ eventually becomes positive, and stays so
sufficiently far in the past to allow $H$ to reach
zero and turn negative. The scale factor $a(t)$ reaches a minimum and 
re-expands again toward the past. This cosmological bounce obtains 
for the spatially closed case ($k=+1$). 

In Figs.(1-3), we confirm numerically the above analysis,
focusing on the latter nonsingular case. (More output 
from the first two cases can be found in [25].)
We simulate that 
scenario starting from the late inflationary stage,
back through the change of signs in ${\dot H}$, and finally, safely
through the cosmological bounce itself. Fig.(1) shows the evolution
of the scalar field $\phi$ in the neighborhood of the bounce. Fig.(2) displays
the overall behavior of the expansion rate $H$, including the inflationary
phases (straight portions of the curve, see Eq.(12)) and the non-inflationary,
or inter-inflationary phases where ${\dot H}$ turns positive and the 
null energy condition is violated. Fig.(3) shows clearly the distinction
between inflationary and inter-inflationary evolutions of the scale
factor, and also confirms that the model produces enough expansion
to be observationally viable. To facilitate the comparison with
the published literature, the figures shown here are obtained 
for $\xi =1000$ and $\lambda = 0.001$, just as in [32,33,35]. Qualitatively,
the results hold for a broad range of parameter values. One can also
obtain some potentially interesting variations by adding a mass term to
the Lagrangian or by specializing to very small or very large parameter
values [25]. Here, our aim is to show that a generic case can be nonsingular.

We have also checked the special case of minimal coupling ($\xi = 0$).
One major difference in this case is that the null energy condition can never
be violated (see Eqs.(4,5)), so that most of the singularity theorems do apply. The sub-cases $k=0$ and $k=-1$ are always singular, 
as the theorems predict [17-21]. For $k=+1$, the situation is less trivial, 
because technically the equations of motion do not exclude a bounce.
(Note that the inflation-specific singularity theorems are also
more technically involved in this case.) 
For instance, one can set $H={\dot\phi}=0$ with $\xi =0$ and $k=1$
in Eqs.(9-11), and successfully evolve the system out of that state of minimal
size in both directions of time. Unfortunately, because of Eq.(4), 
$H$ turns negative almost immediately, and the model recollapses far
too soon to constitute a viable inflationary scenario. One could still
obtain a longer period of expansion by forcing ${\dot\phi}$
to remain virtually zero for a long enough time. For example, this could
be arranged by using a potential that is extremely flat near $\phi =0$
or that has a local minimum around that value, and placing the initial 
$\phi$ and ${\dot\phi}$ very close to zero. But one would then be resorting
to an inflation of the ``new'' or the ``old'' type, respectively. Compared
to chaotic inflation, those two types
have many additional observational and conceptual difficulties 
and are considered today to be among the least attractive [10,41-43].
  
Still on the topic of minimal coupling, 
a very recent work was just brought to our attention which
explores the energy constraints that an inflationary model must satisfy
in order to allow for a cosmological bounce [44]. Unfortunately, that study
misses the possibility (probably the necessity [26-28]) of nonminimal
coupling (e.g., stating inaccurately that all inflationary models are 
minimally coupled.) The study also
concludes that only ``old'' and ``new'' inflation 
(``and not chaotic or eternal inflation'') are compatible with the bounce
scenario from an energetics point-of-view, which is also inaccurate as
we saw above. Finally, reference [44] does not actually work out a specific inflationary model.
Doing so would be necessary because, as we indicated above, one can easily 
construct solutions that look like they could bounce,
either because of their asymptotic behavior or on energetics grounds, but
that are not viable cosmologies [7-9,38,25]. Still, [44] is an interesting extension of previous investigations of the interplay between the energy conditions, the singularity theorems, and inflation.

Our results suggest the possibility of a Universe that inflates, then
reheats (when $\phi$ drops close enough to zero) and expands as a 
Friedmann universe, then recontracts and reheats again, becomes
eventually scalar-field dominated, then bounces when it reaches
a relatively large minimum size, then inflates and reheats again, and so on
for ever in both directions of time (see Figs.(2,3)).
Of course, one would have to contend then with the rapid growth of 
anisotropies and inhomogeneities during contracting phases, and 
with whether black hole formation can be kept in check, e.g., by 
the possible dominance of some type of dark matter, in particular some weakly-interacting, nonbaryonic dark matter. Alternatively, it
 is perhaps possible to cast these findings in the context of 
``eternal'' inflation [11-16], and ask whether the latter can thus
be made eternal in the past as well as in the future [21].

We hope to have shown that it might not be necessary to depart much
from general relativity to avoid cosmological singularities, were they
to be eventually deemed physically or philosophically unacceptable.
If, as it is suggested here, the minimal size to which the Universe recollapses is not too small,
it might not be necessary to call upon quantum gravity in order to 
build fully consistent cosmologies.

During the course of this research, I have benefited from many instructive
exchanges on related technical and philosophical aspects with
R. Brandenberger, V. Husain, E. Luft, S. Savitt, K. Schleich, W. Unruh, P. Varniere, A. Vilenkin, S. Weinstein, and D. Witt.

\clearpage

$^{\$}$Email address: fakir@physics.ubc.ca
\newline\newline
[1] R. Penrose, Phys. Rev. Lett. {\bf 14}, 57 (1965).
\newline
[2] S.W. Hawking and G.F.R. Ellis, {\em The Large Scale Structure
of Space-Time} (Cambridge University Press: Cambridge, England, 1973).
\newline
[3] J. Earman, {\em Bangs, Crunches, Whimpers, and Shrieks: Singularities
and Acausalities in Relativistic Spacetimes} (Oxford University Press:
New York City, NY and Oxford, England, 1995).
\newline
[4] V. Mukhanov and R. Brandenberger, Phys. Rev. Lett. {\bf 68}, 1969 (1992).
\newline
[5] R. Brandenberger, R. Easther and J. Maia, gr-qc/9806111.
\newline
[6] P. Kanti, J. Rizos and K. Tamvakis, gr-qc/9806085.
\newline
[7] S.S. Bayin, F.I. Cooperstock and V. Faraoni, Astrophys. J. {\bf 428}, 
439 (1994).
\newline
[8] S. Capozziello, R. de Ritis and A.A. Marino, gr-qc/9808009.
\newline
[9] A. Billyard, A. Coley and J. Ibanez, gr-qc/9807055.
\newline 
[10] For reviews, see S.K Blau and A.H. Guth, in 
{\em 300 years of Gravitation} (Eds. S.W. Hawking and W. Israel.
Cambridge University Press: Cambridge, England, 1987); 
 A.D. Linde, {\em Particle Physics and Inflationary Cosmology} 
(Harwood Academic, Chur Switzerland, 1990).
\newline
[11] A. Vilenkin, Phys. Rev. D {\bf 27}, 2848 (1983).
\newline
[12] A.D. Linde, Phys. Lett. B {\bf 175}, 395 (1986).
\newline
[13] M. Aryal and A. Vilenkin, Phys. Lett. B {\bf 199}, 351 (1987).
\newline
[14] A.S. Goncharov, A.D. Linde and V.F. Mukhanov, Int. J. Mod. Phys. A 
{\bf 2}, 561 (1987).
\newline
[15] K. Nakao, Y. Nambu and M. Sasaki, Prog. Theor. Phys. {\bf 80}, 
1041 (1988).
\newline
[16] A. Linde, D. Linde and Mezhlumian, Phys. Rev. D {\bf 49}, 1783 (1994).
\newline
[17] A. Borde and A. Vilenkin, Phys. Rev. Lett. {\bf 72}, 3305 (1994).
\newline
[18] A. Borde and A. Vilenkin, in {\em Relativistic Astrophysics: The 
Proceedings of the Eighth Yukawa Symposium} (Ed. M. Sazaki. 
Universal Academy Press: Japan, 1995).
\newline
[19] A. Borde, Phys. Rev. D {\bf 50}, 3692 (1994).
\newline
[20] A. Borde and A. Vilenkin, Int. J. Mod. Phys. D {\bf 5}, 813 (1996).
\newline
[21] A. Borde and A. Vilenkin, Phys. Rev D {\bf 56}, 717 (1997).
\newline
[22] J.D. Bekenstein, Phys. Rev D {\bf 11}, 2072 (1975).
\newline
[23] L. Parker and S.A. Fulling, Phys. Rev. D {\bf 7}, 2357 (1973).
\newline
[24] L. Parker and Y. Wang, Phys. Rev. D {\bf 42}, 1877 (1990).
\newline
[25] R. Fakir, in preparation.
\newline
[26] A.D. Linde, Phys. Lett. B {\bf 114}, 431 (1982).
\newline
[27] D. Freedman, I. Muzinich and E. Weinberg, Ann. Phys. 
{\bf 87}, 95 (1974).
\newline
[28] D. Freedman and E. Weinberg, Ann. Phys. {\bf 87}, 354 (1974).
\newline
[29] V. Faraoni, Phys. Rev. D {\bf 53}, 6813 (1996).
\newline
[30] A. Linde, Phys. Lett. B {\bf 202}, 194 (1988).
\newline
[31] T. Futamase and K. Maeda, Phys. Rev. D {\bf 39}, 399 (1989).
\newline
[32] R. Fakir, {\em Creation of Universes with Nonminimal Coupling}
(Ph.D. thesis, University of British Columbia, Vancouver, BC, Canada, 1989.)
\newline
[33] R. Fakir, Phys. Rev. D {\bf 41}, 3012 (1990).
\newline
[34] R. Fakir and S. Habib, Mod. Phys. Lett. A {\bf 8}, 2827 (1993).
\newline
[35] R. Fakir and W.G. Unruh, Phys. Rev. D {\bf 41}, 1783 (1990).
\newline
[36] R. Fakir, S. Habib and W.G. Unruh, Astrophys. J. {\bf 394}, 396 (1992).
\newline
[37] D. Salopek, Phys. Rev. D {\bf 45}, 1139 (1992).
\newline
[38] R. Brustein and G. Veneziano, Phys. Lett. B {\bf 329}, 429 (1994).
\newline
[39] N. Kaloper, R. Madden and K.A. Olive, Nucl. Phys. B {\bf 452}, 
677 (1995).
\newline
[40] R. Easther, K. Maeda and D. Wands, Phys. Rev. D {\bf 53}, 4247 (1996).
\newline
[41] A.H. Guth and S.-Y. Pi, Phys. Rev. Lett. {\bf 49}, 1110 (1982).
\newline
[42] S. Hawking, Phys. Lett. B {\bf 115}, 295 (1982).
\newline
[43] W.G. Unruh, in {\em Critical Dialogues in Cosmology} 
(Ed. N. Turok. World Scientific, Singapore, 1997).
\newline
[44] C. Molina-Paris and M. Visser, gr-qc/9810023.
\newline

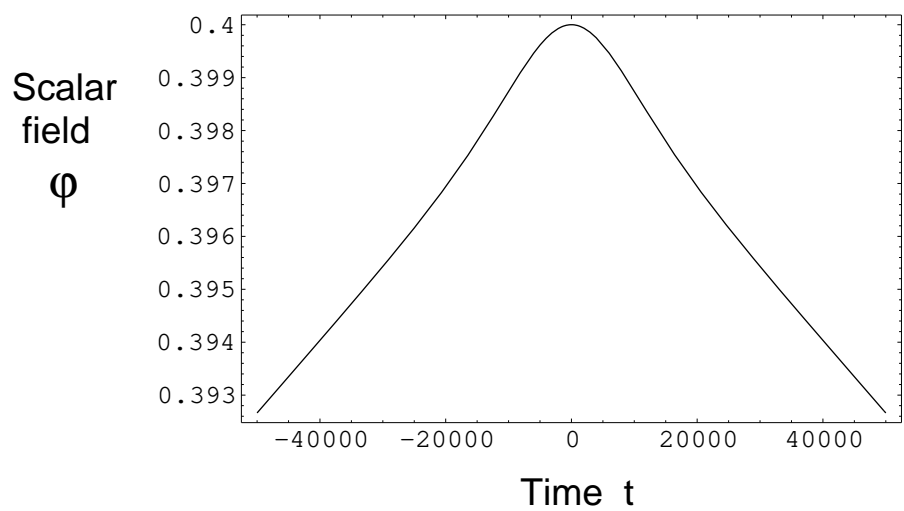
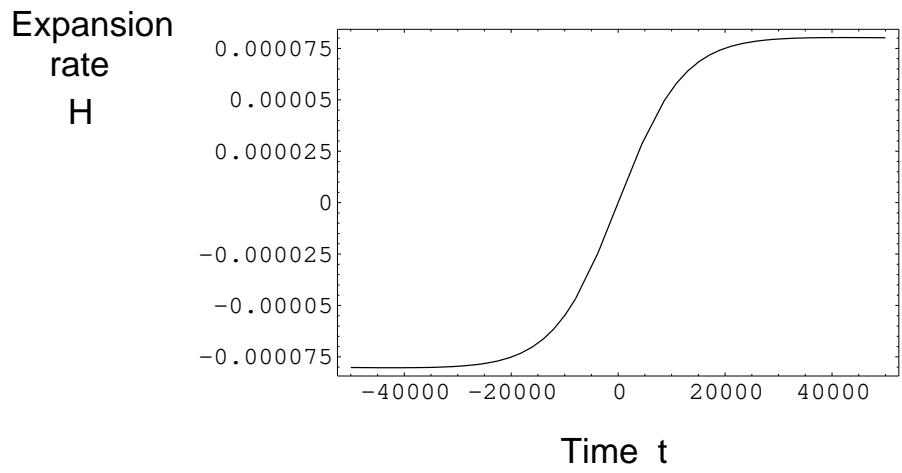
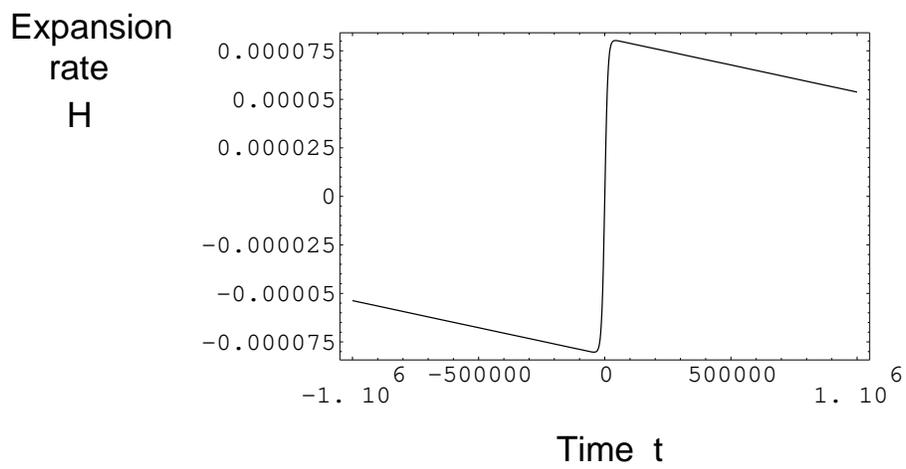
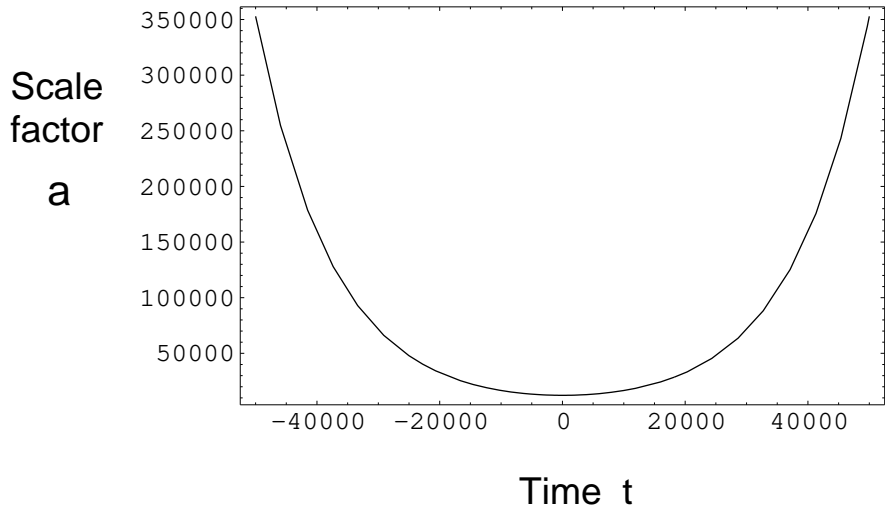
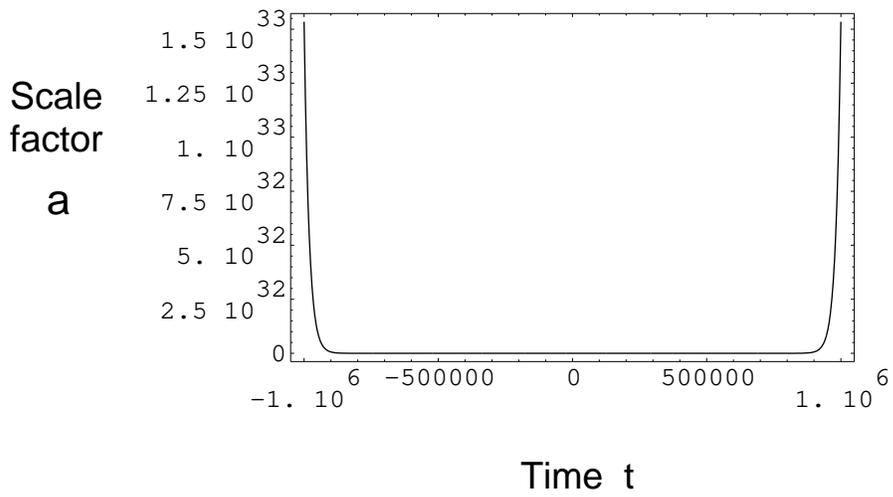
\end{document}